\documentstyle[preprint,aps]{revtex}

\begin{document}

\draft

\title{Addendum: Chaos around a H\'enon-Heiles-Inspired
 Exact Perturbation of a Black Hole}

\author{Werner M. Vieira\thanks{e-mail: vieira@ime.unicamp.br} and
Patricio S. Letelier\thanks{e-mail: letelier@ime.unicamp.br}}

\address{Departamento de Matem\'atica Aplicada\\
Instituto de Matem\'atica, Estat\'{\i}stica e Ci\^encias da
 Computa\c{c}\~ao\\
Universidade Estadual de Campinas, CP 6065\\
13081-970 Campinas, SP, Brazil}

\date{August 29, 1996}

\maketitle

\begin{abstract}

We find that the model of a black hole plus an exterior halo
of quadrupoles and octopoles recently proposed by us
is more chaotic than previously detected.
In fact, the quadrupolar component gives rise also to a
chaotic behavior, found after further numerical search.
This fact reinforces even more the role of chaos in relativistic
core--halo models.\\

\end{abstract}


We have recently proposed~\cite{WP} an
exact relativistic model to describe core-halo type gravitational
 systems.
 The astronomical motivation for the model was a massive black hole
 or a galactic nucleus at the center, surrounded by
an also massive, exterior halo of dust.
 This halo does not need to be planar and was inspired
 by the classical 
H\'enon-Heiles system in the Newtonian limit of
 General Relativity.
 To this end, we 
retained both the quadrupolar and octopolar terms in the
 halo expansion.
The core--halo solution was constructed in Weyl
 metric, where a kind
 of nonlinear superposition of distinct solutions is known to be
 possible.

We also studied in~\cite{WP} the influence of the exterior halo
 on the geodesic motion in the inner vacuum. We detected the
 occurrence of large chaotic regions due to the presence of
 the octopolar halo component (see Fig.\ 1(b) of that Reference)
 and an apparent preservation of integrability when only
 the quadrupolar halo component was taken into account 
(see Fig.\ 1(c) there).

In this note we report the occurrence of chaos even when we
 have a purely quadrupolar halo. In this case,
 the chaotic regions are much smaller than in the octopolar
 case, being found after a more detailed numerical search.
 This is shown in Fig.\ 1, which is to be compared with
 Fig.\ 1(c) of~\cite{WP}. It stresses the fact that chaos
 is more conspicuous in our model than previously detected
 and hence reinforces its role in relativistic core--halo
 models. On the other hand, this additional numerical finding
 does not alter any other conclusion of~\cite{WP}; in particular,
 the presence of both the quadrupolar and the octopolar terms
 are indispensable to accomplish the H\'enon-Heiles--like form
 of the halo in the Newtonian limit.

The additional finding above recalls us of a basic question:
 how much
 information is available about a system from Poincar\'e's
 sections? Although powerful to identify
in an invariant way
chaotic signatures in phase space,
 we know that it is a
 numeric tool and hence it is insufficient in any case to assure
 full integrability.

In~\cite{Moeckel} a perturbation of the Schwarzschild
background due to an exterior halo of matter
is also considered. It reduces
in the Newtonian limit to the quadrupolar
gravitational moment (instead of a dipolar one as found
in~\cite{WP}). In~\cite{Cornish} such a multipolar
expansions are also considered for Majumdar--Papapetrou
solutions. In constrast with this, each one of our
$n^{th}$--pole, besides being {\it{per se}}
an exact solution, it is a pure one in the sense
that it leads to
the corresponding Newtonian pole in that limit.
Finally, Moeckel's claim that his model is the relativistic
analog of the famous Hill problem in celestial mechanics
is unclear and will be considered in another work.

The authors thank  CNPq and FAPESP for financial support.

\begin{figure}

\caption{This figure is in all respects the same
as Fig.\ 1(c) of \protect\cite{WP},
except that it contains much more points,
this time sufficient to 
exhibit (three small) chaotic regions (one
around each vertex of the triangular region).}
\end{figure}


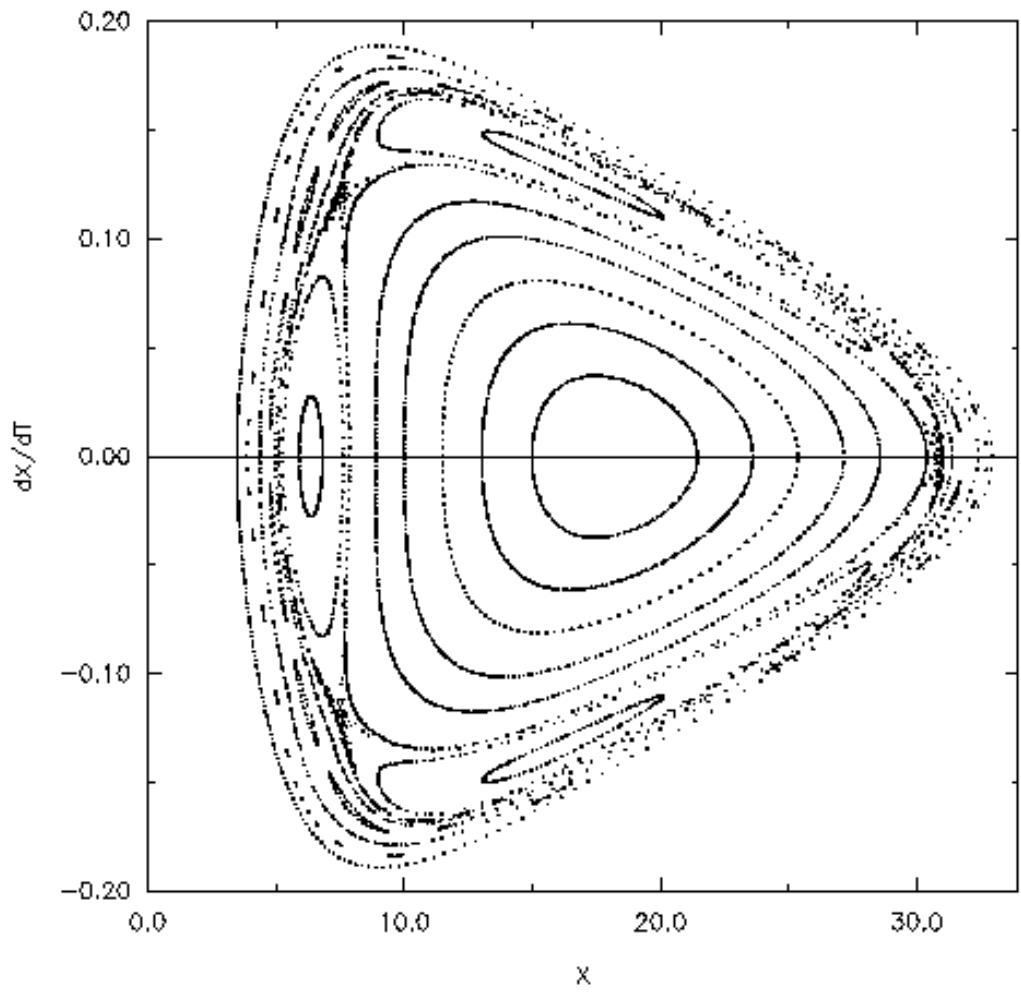
\end{document}